\documentclass[12pt,a4paper,one column]{article}

\usepackage[utf8]{inputenc}
\usepackage{amsmath}
\usepackage{amsfonts}
\usepackage{amssymb}
\usepackage{graphicx}
\usepackage{graphics}
\usepackage{color}
\title{Study of complex properties of binary system of ethanol-methanol at extreme concentrations}
\author{K Nilavarasi, Thejus R Kartha and V Madhurima*\\
Department of Physics\\
School of Basic and Applied Sciences\\
Central University of Tamil Nadu\\
Thiruvarur - 610101\\
*madhurima@cutn.ac.in
}
\date { }

\begin{document}
\maketitle
\begin{abstract}
At low concentrations of methanol in ethanol-methanol binary system, the molecular interactions are seen to be uniquely complex. It is observed that the ethanol aggregates are not strictly hydrogen-bonded complexes; dispersion forces also play a dominant role in the self-association of ethanol molecules.  On the addition of small amount of methanol to ethanol, the dipolar association of ethanol is destroyed. The repulsive forces between the two moieties dominate the behavior of the binary system at lower concentration of methanol. At higher concentration of methanol $(>30\%)$, the strength and extent (number) of formation of hydrogen bonds between ethanol and methanol increases. The geometry of molecular structure at high concentration favors the fitting of component molecules with each other. Intermolecular interactions in the ethanol-methanol binary system over the entire concentration range were investigated in detail using broadband dielectric spectroscopy, FTIR, surface tension and refractive index studies. Molecular Dynamics simulations show that the hydrogen bond density is a direct function of the number of methanol molecules present, as the ethanol aggregates are not strictly hydrogen-bond constructed which is in agreement with the experimental results.

\end{abstract}



\section{Introduction}
\label{}
Ethanol and methanol strongly self-associate due to hydrogen bonding \cite{markus(1998),funel(1994),saiz(1997),wertz(1967)}. These primary polar liquids with C-OH group form very similar hydrogen bond acceptors and hydrogen bond donors and they exhibit extended network \cite{wu(2007)}. On mixing, these two liquids does not form azeotropes.  Also, methanol does not form azeotrope with water, whereas ethanol forms an azeotrope with it. Although the binary system of ethanol and methanol is assumed to be ideal, their behavior at low concentrations is far from ideal \cite{amer(1955)}.\\  

Vast experimental information is available concerning the dielectric properties, excess volume parameters, FTIR spectra and refractive index of binary system of ethanol-water \cite{per(1969)},methanol-water \cite{per(1969),xiao(1997)}, ethanol-higher alcohols \cite{jose(2015)}, methanol-higher alcohols \cite{lone(2011)}, ethanol-pyrindine \cite{ezekial(2012)}, methanol-pyridine \cite{ezekial(2012)}, etc. All these systems show a deviation from ideal behaviour for most physical and chemical properties at $50\%$. On the other hand, comparatively very few reports are available for the ethanol-methanol binary system.  Amer et. al., found the non-ideal behavior of ethanol-methanol at low concentrations while investigating the activity measurements of ethanol-methanol-acetone system \cite{amer(1955)}.  The current study is undertaken to understand the anomalous behavior of intermolecular interactions of ethanol-methanol system.  Dielectric, FTIR, surface tension, density, refractive index and molecular dynamics studies for this system is undertaken over the complete concentration range, with emphasis on concentration regions where anomalous behavior in properties is seen. The excess values of molar volume and refractive index, molar refraction and total partial pressure were determined using the experimental values. Complex impedance Cole-Cole plots were used to study the relaxation mechanism in the ethanol-methanol system. \\

Molecular dynamics simulations were done to verify the behavior of ethanol-methanol binary system. The structure of molecules and the number of hydrogen-bonded molecules determined from the simulations are used to interpret the the experimental results of surface tension, density, refractive index and dielectric values relating hydrogen bond structure. It is understood from the present study that the methanol molecules are responsible for increase in hydrogen bond density and they play the role of a mediator in connecting ethanol molecules. 

\section{Experimental section}
\subsection{Materials}
Methanol and ethanol were purchased from Sigma Aldrich and Merck Emplura respectively.  The purities of methanol and ethanol were $99.9\%$. The precision of the binary system was $\pm 0.1 mg$ and they were measured using a weighing balance.  
\subsection{Methods}
Density of the binary systems were measured by using a $25 ml$ specific gravity bottle, calibrated with distilled water.  Surface tension of the binary systems over the whole concentration range were measured using Rame-Hart contact angle goniometer.  Pendant drop method was used in calculating the surface tension. The polar and dispersive parts of surface tension were calculated using the contact angle datas.  The mid infra-red spectra of the binary systems were measured by Fourier Transform Infra Red Spectrometer of Perkin Elmer.  Broadband dielectric studies were carried out using Vector Network Analyzer with Dielectric Assessment Kit of Rhode and Scharwz. The refractive index measurements with Na light were carried out for the binary system using Abbe refractometer which was calibrated using doubly distilled water. All the experiments were performed at room temperature. 

\subsection{Computational details}
The well-known structures of methanol and ethanol were first constructed using the Avogadro package \cite{hanwell(2012)}. The saved input files were optimized using the GAMESS \cite{schmidt(1993)} package, using an ab-initio Hartree-Fock \cite{root(1951), szabo(1996)} with a $6-31G^{*}$ basis set \cite{franc(1982), hari(1973)}.  The simulations were carried out using GROMACS \cite{becker(1993), berendsen(1995), lindahl(2001), van(2005), hess(2008), pronk(2013), pall(2015), abraham(2015)}. Visualizations of the simulations were obtained using the VMD software \cite{humphrey(1996)}.\\

Methanol and ethanol mixtures from $0\%$ to $100\%$ proportion of methanol were produced. Further details of the solution concentrations are given in table. All the samples were subjected to energy minimization procedures, using the steepest descent algorithm \cite{zimmer(1991)}, with an energy tolerance of 500 kJ/mol. Any outlying molecules were manually replaced, assuring that the concentration remain unchanged.\\

Following the energy minimization, the system was subjected to two equilibration processes. First is an NVT equilibration in which the thermostat algorithm is used to set the temperature right. This thermally equilibrated system is then subject to an NPT equilibration wherein the system is compressed to obtain the correct density. This is done using a barostat \cite{parri(1981)} and a thermostat algorithm \cite{beren(1984)}. We can now conclude that the system is very much comparable to a laboratory case. \\

\section{Results and Discussion}
Methanol is a polar molecule capable of H-bonding both with itself and with other oxygen or nitrogen containing molecules, like
water, thus making it completely soluble in water. However, with a single methyl group, methanol has only weak London dispersion forces with itself and with other molecules \cite{markus(1998)}. \\

With an increase in the alkyl chain, ethanol has an increase in the dispersion forces which is comparably greater than the stronger dipole-dipole and hydrogen bonding intermolecular forces in alcohols \cite{saiz(1997)}. As a result, methanol is insoluble in alkanes, such as hexane and heptane, whereas ethanol and propanol are completely miscible with these alkanes.\\  

On the addition of the two liquids, there will be an expansion or contraction in volume of the mixture \cite{sara(2012)}. The volume expansion occurs when the  geometry of molecular structure is such that it does not fit one component of molecules into the other. The steric hindrance and the loss of dipolar association can also lead to expansion in the volume of the mixture. Volume contraction occurs when there is an increase in chemical interaction between the constituent molecules and if the geometry of molecular structure favors the fitting of component molecules with each other, volume of the mixture decreases.  The contraction shows that the molecules of one component are accommodated into the interstitials of the other component\cite{sara(2012)}.\\     

On addition of methanol to ethanol, the hydrogen bonds in ethanol aggregates are disrupted by the methanol molecules. Subsequently when the hydrogen bond strength between ethanol molecules overcomes the dispersive interaction between ethanol and methanol, methanol molecules are encaged by ethanol. On further addition of methanol, bridged structures of ethanol-methanol are formed.\\ 
\begin{figure}[h]
\centerline{\includegraphics[width=5in]{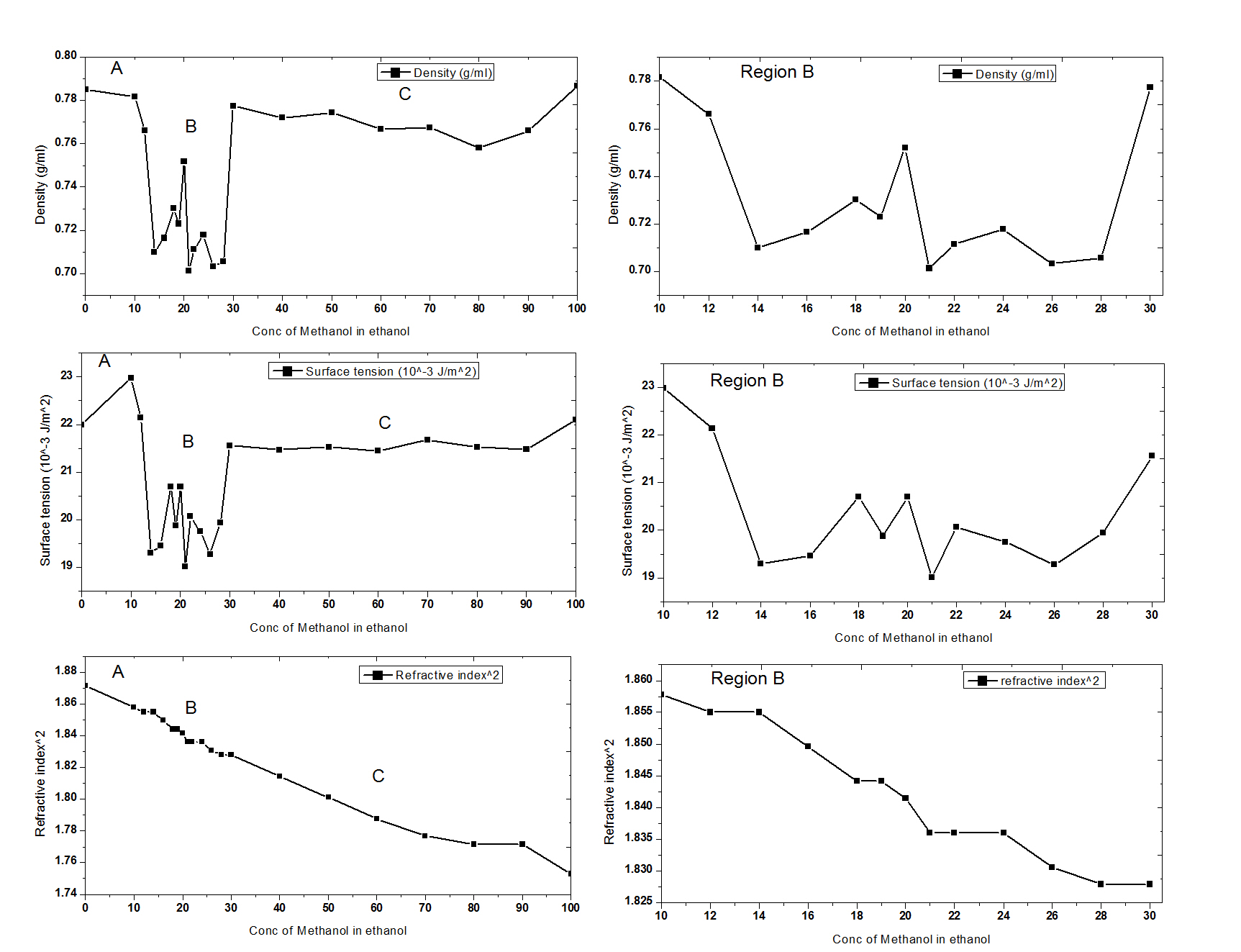}}
\caption[]{Variation of density, surface tension and refractive index with mole fraction of methanol}%
\label{1}
\end{figure}
\subsection{Density and surface tension}
The experimental values of surface tension, density and refractive index plotted against the concentration of methanol are shown in Figure~\ref{1}. In the figure, region A, B and C indicates $<10\%$, $10\%$ to $30\% $ and $>30\%$ of methanol in ethanol respectively.  Usually with the increase in hydrogen bond density, the density of the system will increase. The surface tension is a measure of strength of intermolecular forces and in region A, the surface tension shows a slight increase with an addition of methanol. In region A, density $(\rho)$ is seen to drop dramatically with an addition of small amount of methanol to ethanol.  The decrease in density indicates the decrease in magnitude of volume contraction on mixing of two liquids.  It is also clear that, with the addition of methanol, the dipolar association of ethanol molecules start to break.  \\

On further addition of methanol, the system shows a complex variation in density and surface tension at the region B. The observed increase/decrease in density at this region indicates the increase/decrease in magnitude of contraction on mixing of these liquids.  In particular, a dramatic increase in  contraction is observed at $19\%$ of methanol concentration.  This indicates that at $19\%$ of methanol, the interaction between ethanol and methanol molecules is large and  this leads to accommodation of methanol molecules within ethanol aggregates. At $21\%$, the density drops to 0.70 g/ml indicating the decrease in magnitude of contraction that results from the break up of interaction between molecules by the rupture of hydrogen bonded chains and loosening of dipole interactions.  As a whole, in the region B, dipole-dipole interactions and hydrogen bonding that exists in pure ethanol and pure methanol decrease and the intermolecular interactions between ethanol and methanol increase. This is also confirmed from the surface tension measurements. \\    

Above $30\%$ of methanol (region C), the density and surface tension is constant indicating the equilibrium attained by the system.  It is also clear that with the increase in methanol concentration, the hydrogen bond density increases.  \\  
\begin{figure}[h]
\centerline{\includegraphics[width=5in]{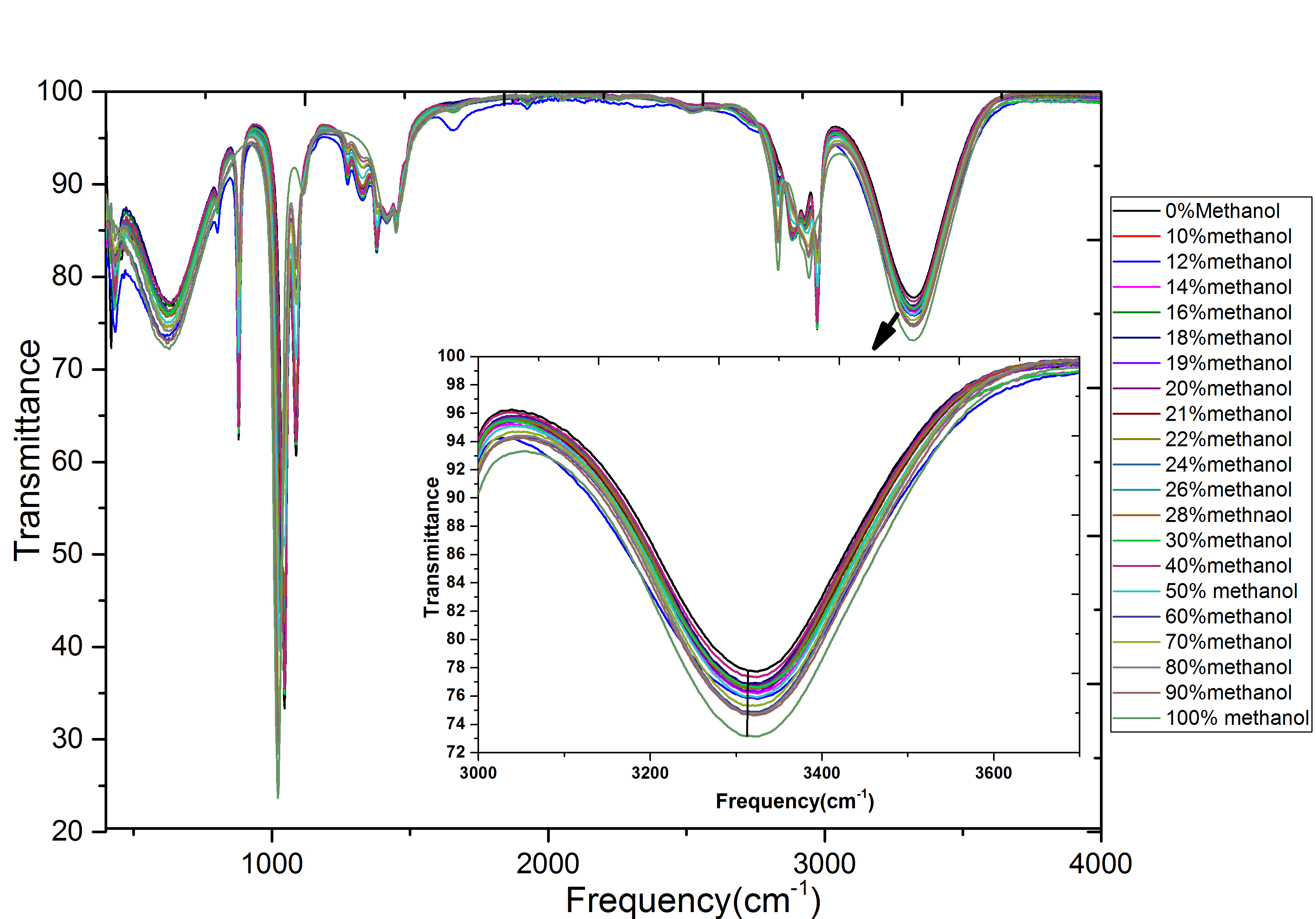}}
\caption[]{The FTIR spectra for various concentration of methanol in ethanol}%
\label{2}
\end{figure}
\subsection{Refractive index}
There is an non-ideal decrease in refractive index of the binary system from 1.361 to 1.327 with decrease in ethanol concentration. From the figure 1, a decrease in refractive index is observed at region A and C that indicates that the polarization due to electrons is reduced with an addition of methanol to ethanol.  \\ 

In the region B (concentration range of $10\%$ to $30\%$), an interesting behavior of refractive index is observed.  It is observed that the refractive index remains unchanged at $12\%$ to $14\%$, $18\%$ to $19\%$ and $21\%$ to $24\%$ of methanol in ethanol.  This is an indication of clathrate formation ie., the formation of open structures containing cages.  The cages are generated due to the hydrogen bonding between the molecules of solute being added.  Here the clathrate formation is mainly attributed to the hydrogen bonding between methanol and ethanol molecules. This also shows that electronic polarization is stable at these concentrations. \\
\begin{figure}[h]
\centerline{\includegraphics[width=5in]{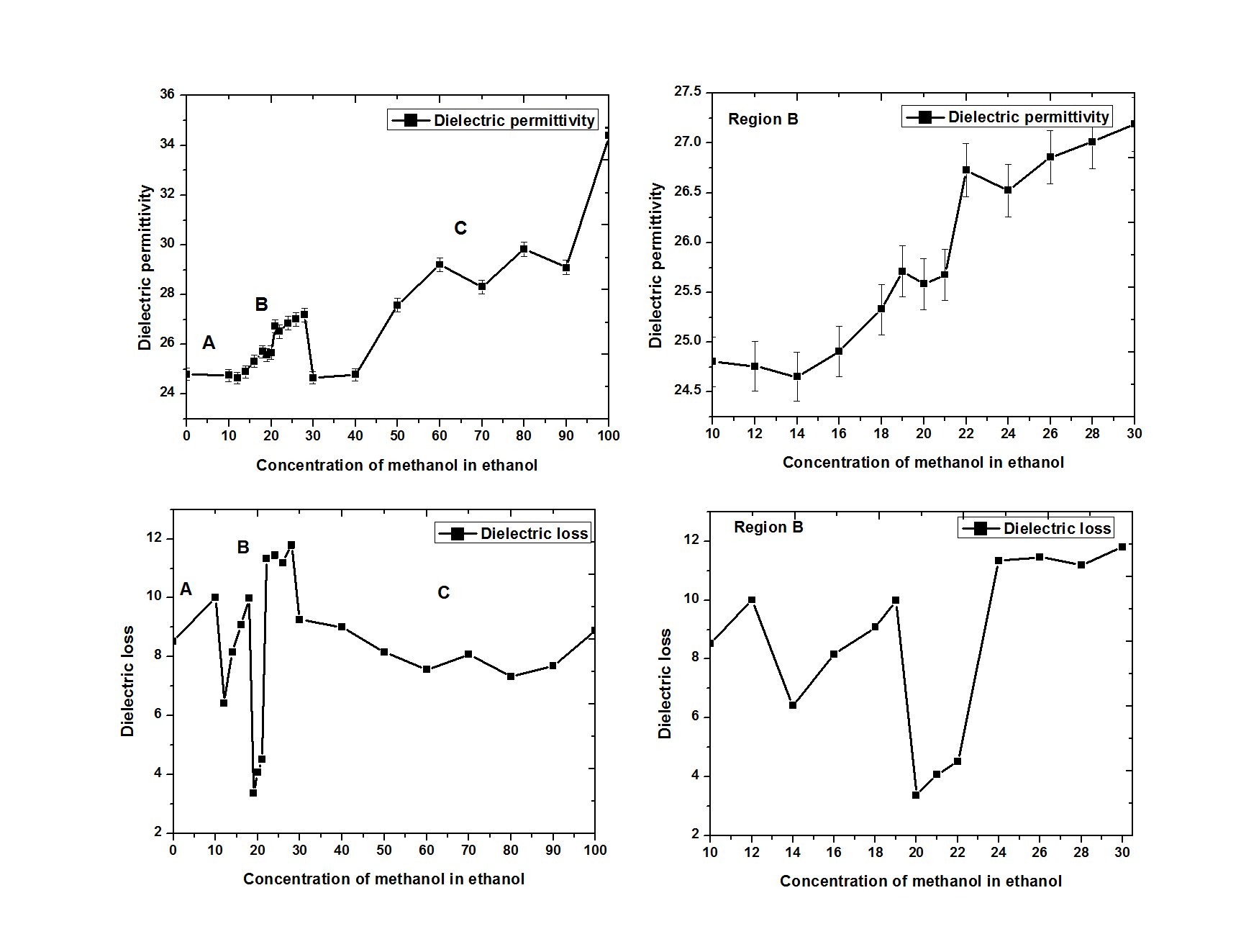}}
\caption[]{Variation of dielectric permittivity with mole fraction of methanol.}%
\label{3}
\end{figure}
\subsection{Dielectric dispersion}
On examining the static dielectric permittivity of the binary system, there is an increase in permittivity with respect to the concentration of methanol as shown in Figure~\ref{3}.  In region A, the binary system doesn't show any variation in permittivity. On further increasing methanol concentration, the static permittivity at region B shows a gradual increase in permittivity till $18\%$, a fall at $19\%$ followed by a sharp rise at $21\%$ of methanol. There is no uniformity in the dielectric values unlike in refractive index.  This change clearly indicates the formation of temporarily induced dipoles and the increase in corresponding dispersion interaction between the ethanol-methanol molecules at region B.  This also confirms the breaking of hydrogen bond network of like molecules and formation of hydrogen bonded ethanol-methanol bridged structures which is also reflected in the FTIR spectra. At region C, the dielectric permittivity is observed to increase with methanol concentration indicating the formation of dipoles.\\  
\begin{figure}[h]
\centerline{\includegraphics[width=5in]{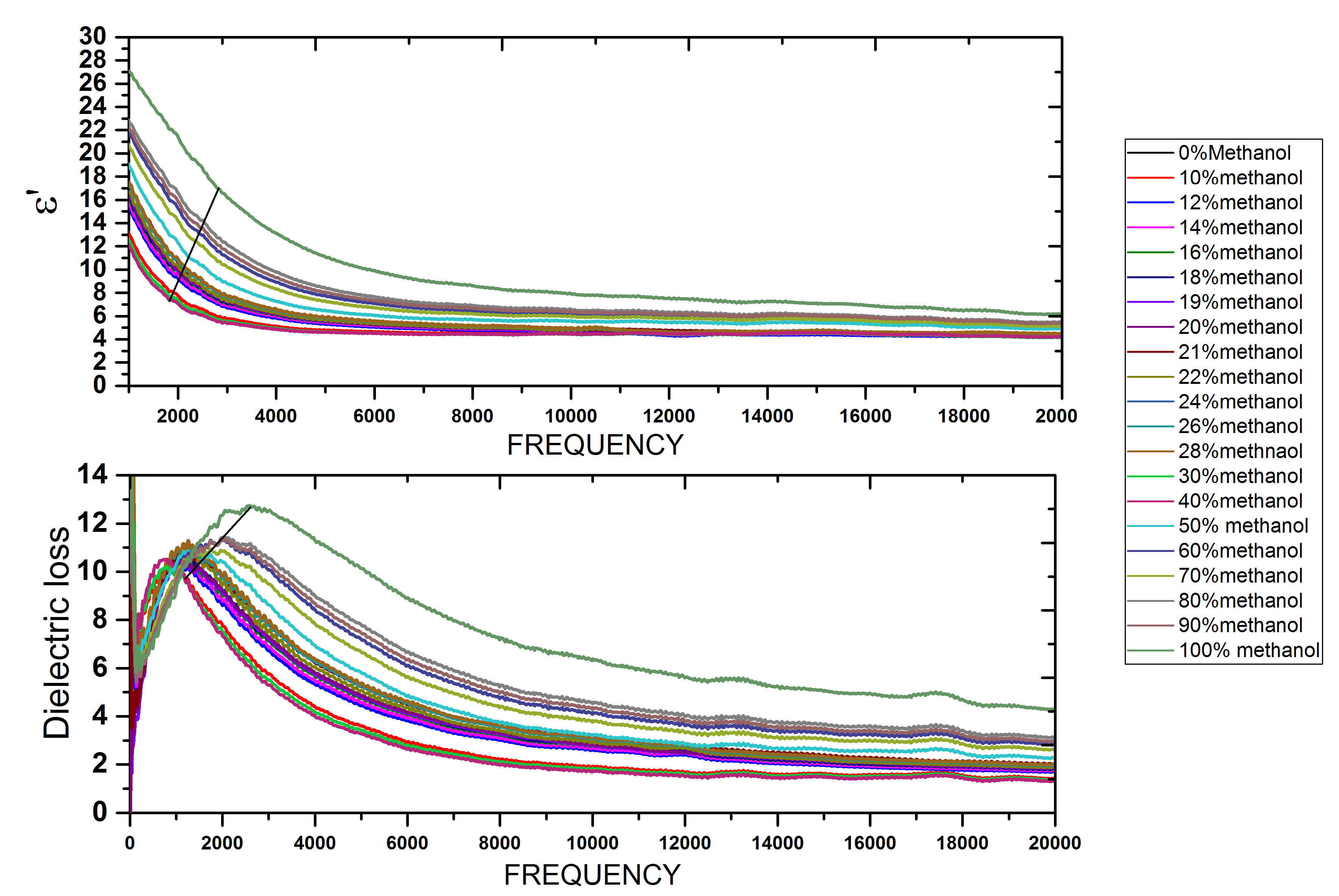}}
\caption[]{Variation of dielectric permittivity  and dielectric loss with frequency for all mole fraction of methanol.}%
\label{4}
\end{figure}
The dielectric loss is seen to drop at region B ($10\%$ to $30\%$ of methanol), which confirms the polarization due to temporarily induced dipoles.  The dielectric loss for all concentrations are shown in Figure~\ref{3}. The dielectric permittivity and dielectric loss of the binary system for the frequency range $20 MHz$ to $20 GHz$ are shown in Figure~\ref{4}.\\
\begin{figure}[h]
\centerline{\includegraphics[width=5in]{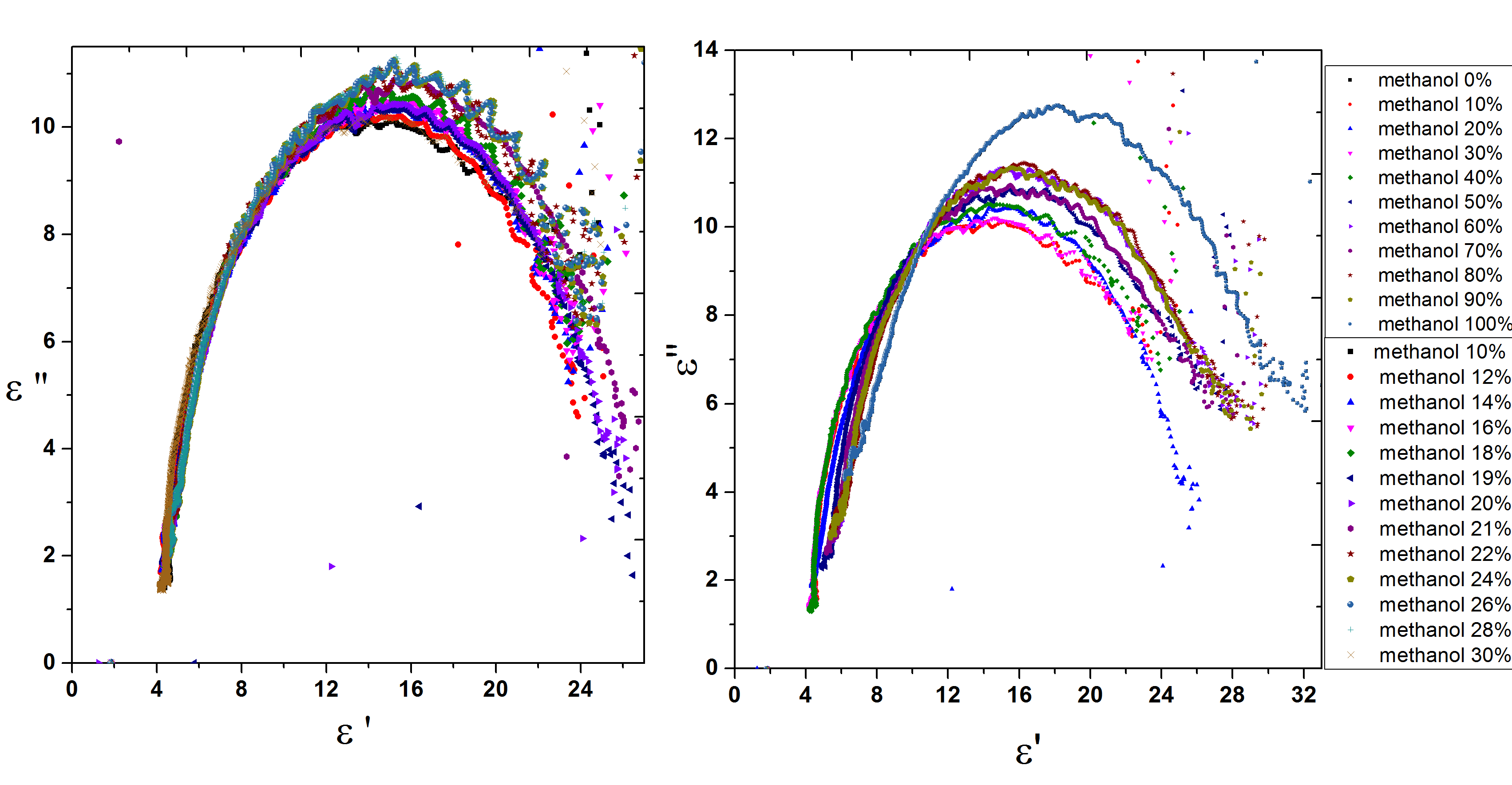}}
\caption[]{Cole-cole plots for various mole fraction of methanol.}%
\label{5}
\end{figure}
The further analysis of measured complex permittivity is performed using the curve fitting technique based on the Cole-Cole model. This is done to find the relaxation time. The Cole-Cole plots of all the concentrations are performed and are shown in Figure~\ref{5} and Figure~\ref{6}. At high frequencies above $12GHz$, the Cole-Cole plot of all concentration of methanol  is located on the same line.  This implies that at all mole fractions of methanol in ethanol, the relaxation mechanism is same in this frequency region.  Permittivity being on the same line at this frequency region indicates that all Cole-Cole plots must have have the same limiting high frequency, $\epsilon_{\infty}$. In this region, the relaxation laws are the same and the same dipole moment is responsible for the observations between $12GHz$ to $20GHz$.  Both ethanol and methanol contain OH group and the dielectric signal observed in this high frequency region is attributed to the dipole moment of the OH groups.  \\

At low frequencies, the Cole-Cole plots have the same shape of varying magnitudes. The relaxation at this low frequency region is associated with the instantaneous dipoles formed and its re-orientation among the like molecules. Deviation of some points from Cole-Cole plots is attributed to the presence of interacting species in various concentrations. \\  
\subsection{FTIR }
The hydrogen bonds in methanol are stronger than in ethanol whereas in ethanol, dispersion forces dominate over hydrogen bonds.  So with the addition of methanol to ethanol, weakening of dispersion forces take place. The breaking up of hydrogen bond network of ethanol and formation of ethanol-methanol hydrogen bonds are seen as alternate red and blue shifts in the OH stretching peak of the infra red spectra.  The red shift  of $8\%$ in OH stretching frequency at $16\%$ of methanol manifests the strength of formation of hydrogen bonds.  The IR spectrum with respect to the molar fraction of methanol is shown in Figure~\ref{2}. The hydrogen bonds start to form between unlike molecules and get stabilized after $30\%$ of methanol concentration in ethanol.  Beyond $30\%$, as expected, the number of hydrogen bonds between ethanol and methanol molecules increase and is shown with a red shift in the transmittance spectra.\\   

\begin{figure}[h]
\centerline{\includegraphics[width=5in]{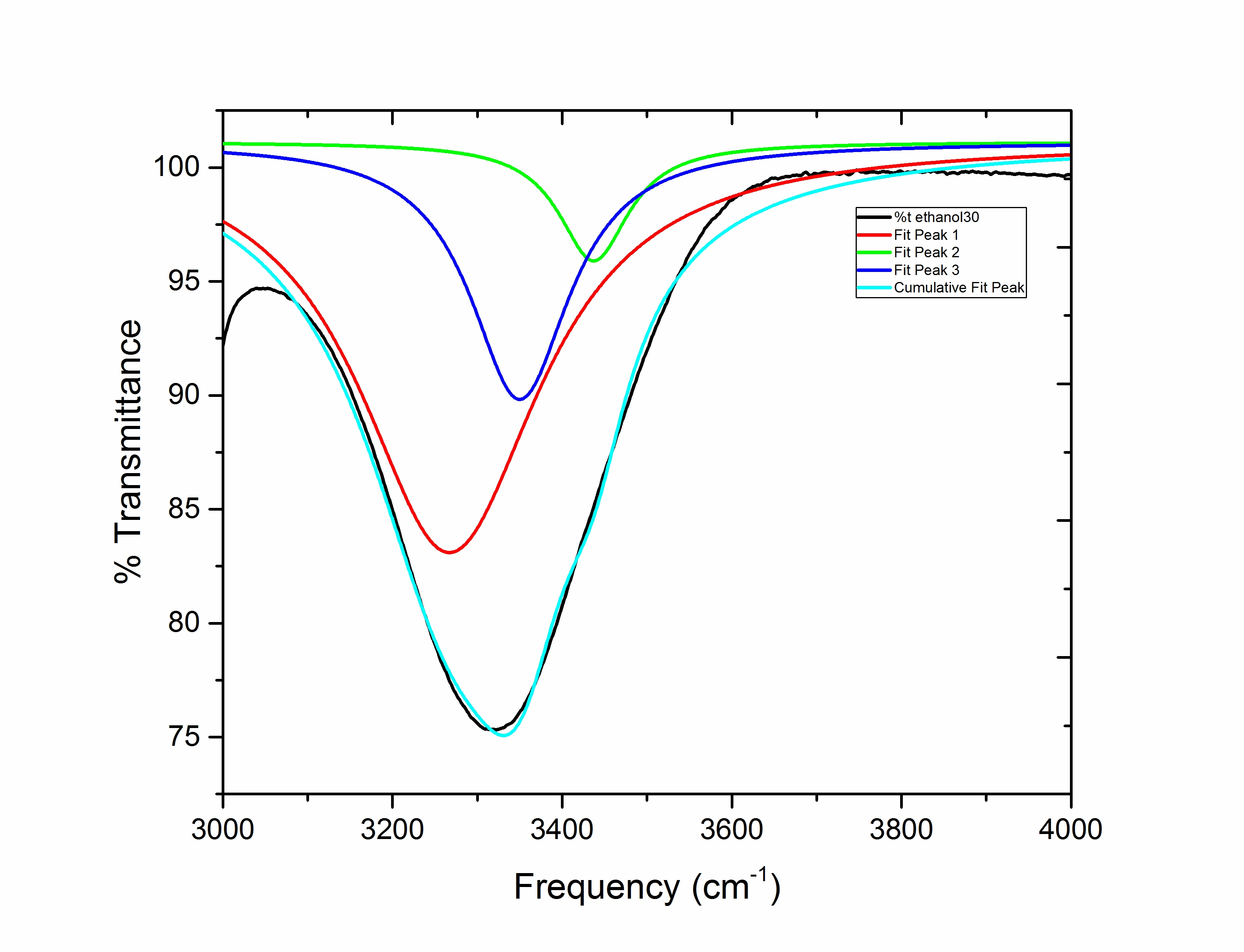}}
\caption[]{Deconvolution of O-H peak for $21\%$ mole fraction of methanol.}%
\label{6}
\end{figure}

On deconvultion of the OH peak, three peaks of frequency $3350 cm^{-1}$, $3433 cm^{-1}$ and $3280 cm^{-1}$ corresponding to multimers, open chain tetramers and cyclic multimers are found.  The deconvoluted spectrum for the concentration of $21\%$ of methanol is shown in Figure~\ref{6}.\\ 

\begin{figure}[h]
\centerline{\includegraphics[width=5in]{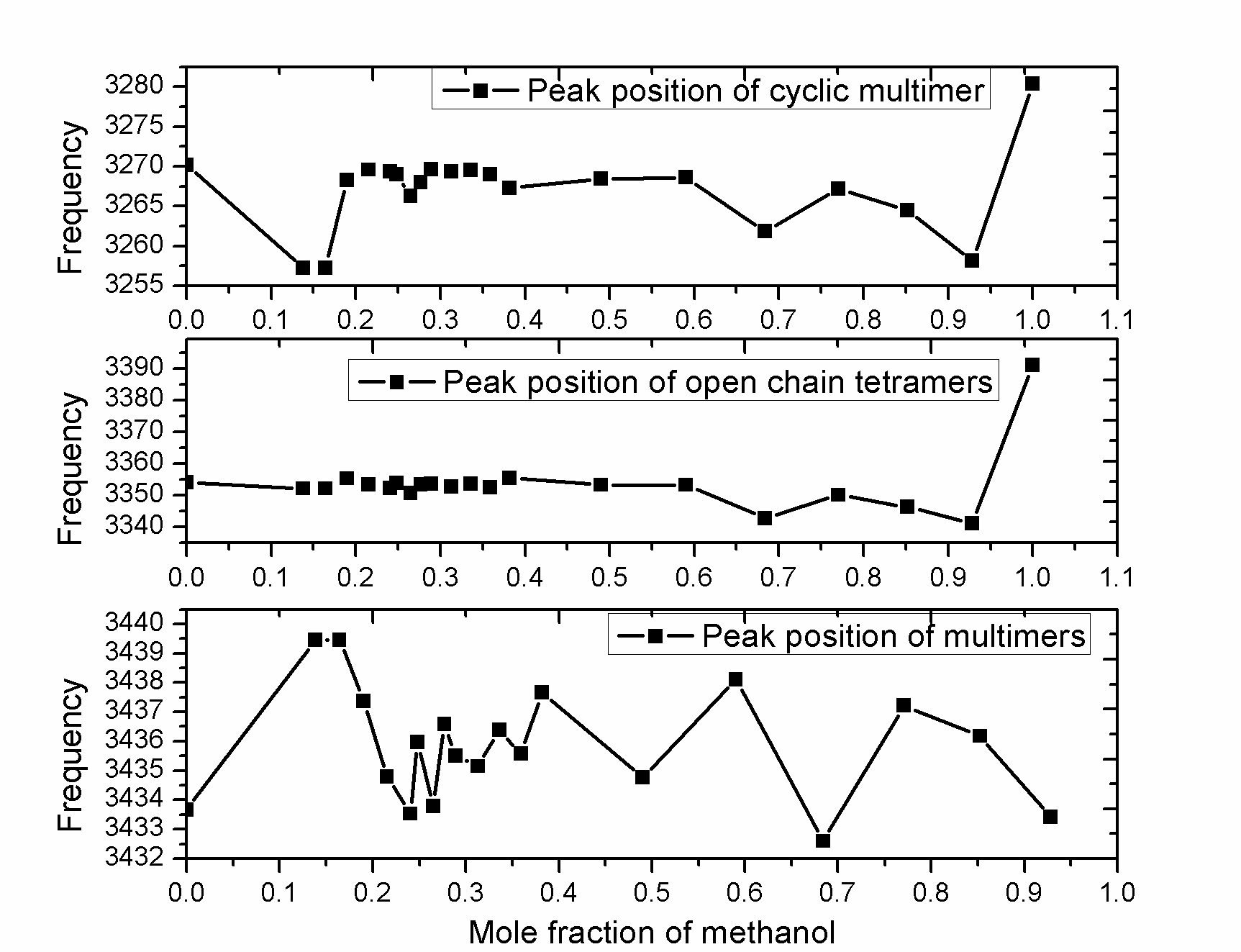}}
\caption[]{variation of peak position of deconvoluted O-H peak with mole fraction of methanol.}%
\label{7}
\end{figure}
\begin{figure}[h]
\centerline{\includegraphics[width=5in]{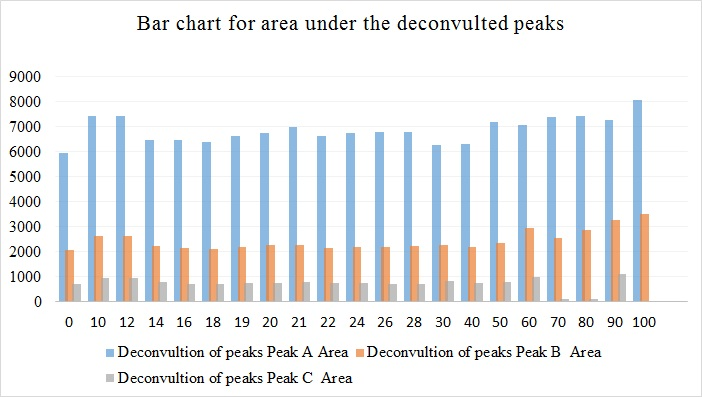}}
\caption[]{Area of deconvuluted O-H peaks  with mole fraction of methanol.}%
\label{8}
\end{figure}

The deconvulted OH peak positions are shown in Figure~\ref{7}. From the figure it is observed that the multimer peak shows alternate red and blue shifts.  This confirms the instability of intermolecular interactions with the increase in concentration of methanol.  This also confirms the formation of long chain multimers composed of ethanol-methanol and ethanol molecules.  The position of the peak corresponding to the cyclic multimers seem to be at the same frequency across the concentration range of $10\%$ to $30\%$ of methanol in ethanol(region B). This attributes to clathrate formation as seen from the refractive index measurements. The comparison of the area under the three deconvoluted peaks is shown in the Figure~\ref{8}.  It is observed from the figure that, the formation of cyclic multimers are more compared to tetramers and pentamers.\\
\subsection{Excess molar volume}
\begin{figure}[h]
\centerline{\includegraphics[width=5in]{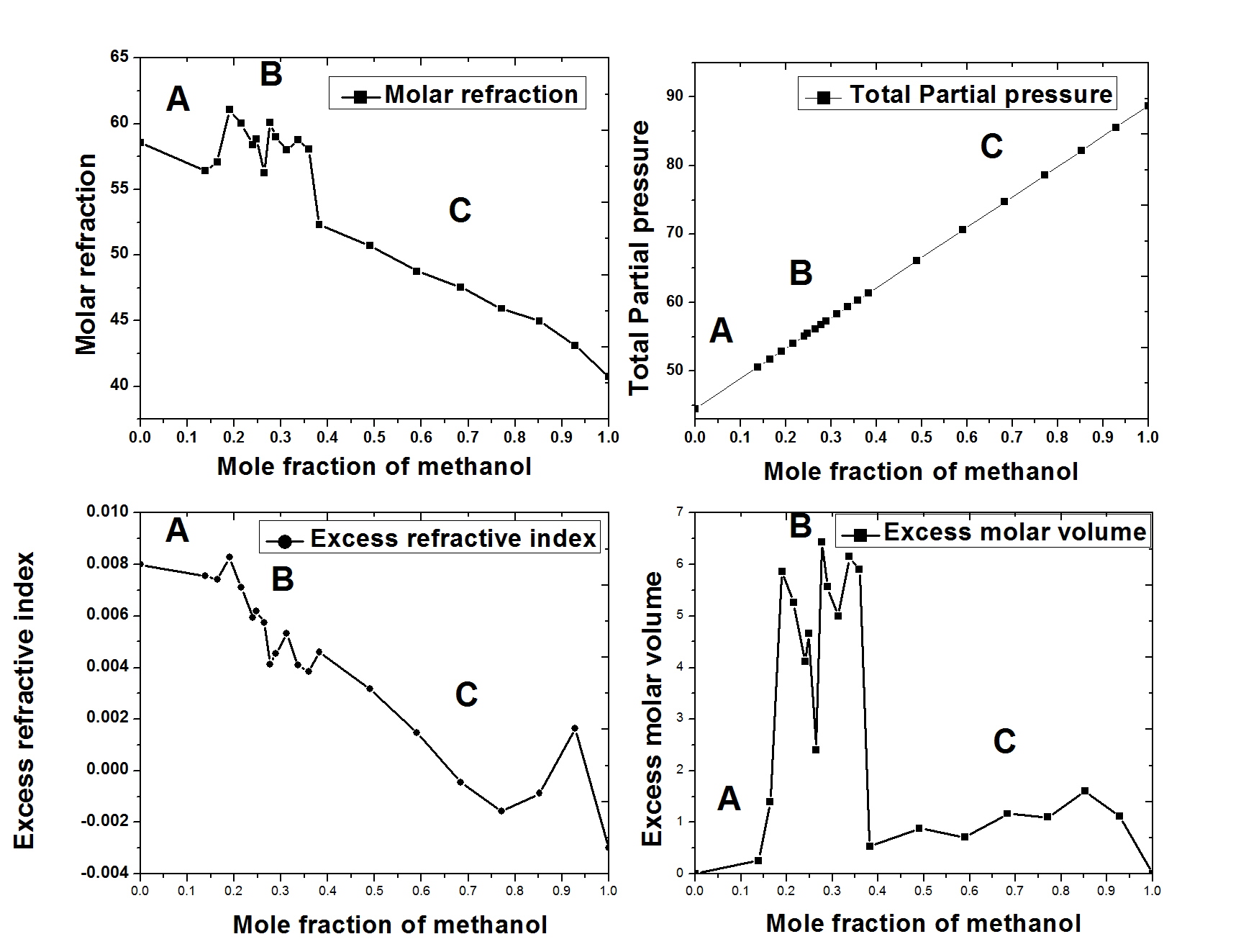}}
\caption[]{Variation of excess values of molar volume, molar refraction, refractive index and total partial pressure with mole fraction of methanol.}%
\label{9}
\end{figure}

\begin{figure}[h]
\centerline{\includegraphics[width=5in]{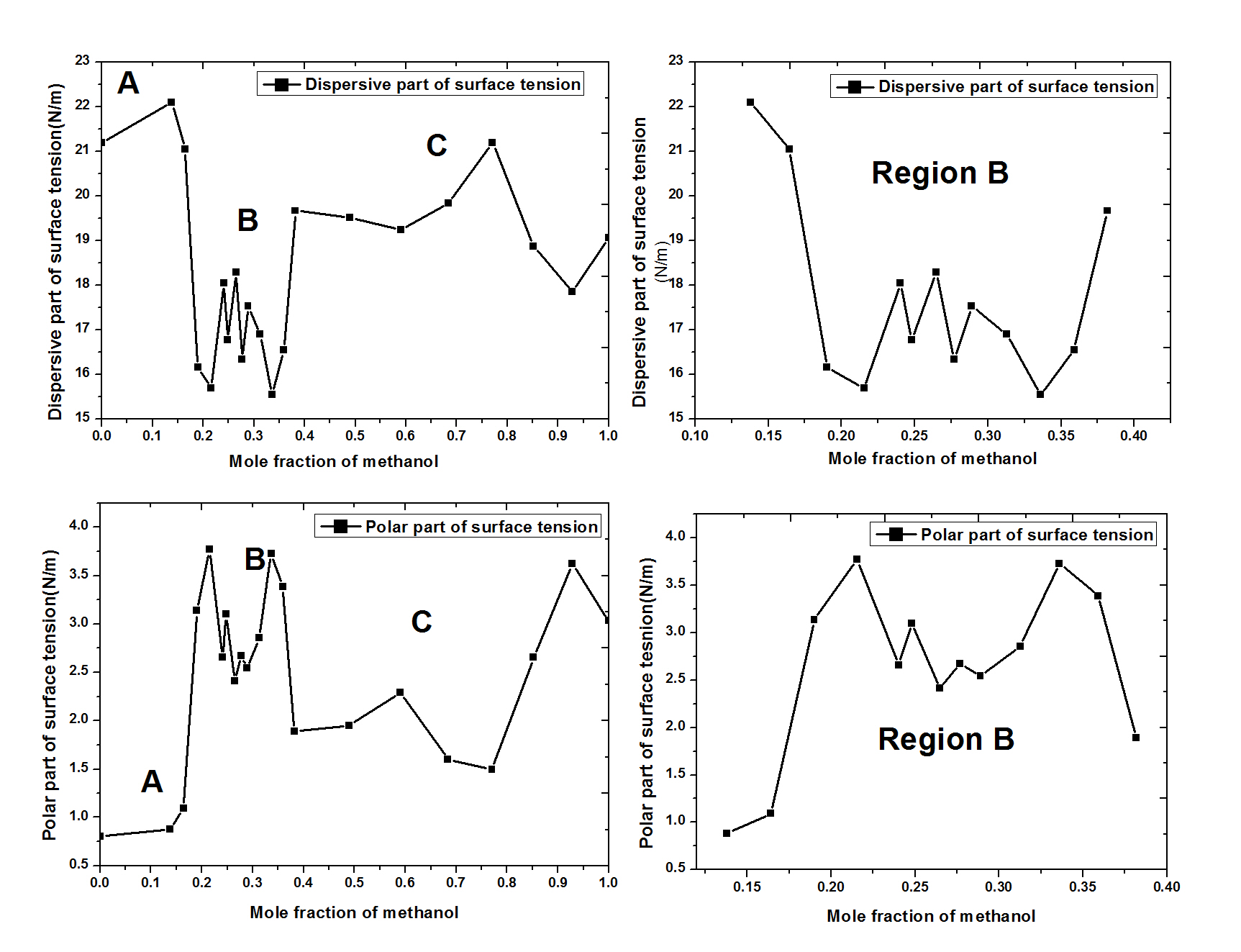}}
\caption[]{variation of polar and dispersive parts of surface tension with mole fraction of methanol.}%
\label{10}
\end{figure}

To further confirm the role of dispersion forces in the binary system, the excess values of properties are calculated and found that this binary system deviates from its ideal behavior [Figure~\ref{9}].  From the excess molar volume plot, it is clear that volume increases for the concentration of $10\%$ to $30\%$.  This confirms that the strength of intermolecular forces decreases. This is attributed to the  formation of multimer structures through dispersion forces between the ethanol molecules, along with a hydrogen bond formed between ethanol and methanol. \\ 
This dominant role of dispersive forces is further confirmed by measuring the polar and dispersive parts of surface tension using the contact angle values.  Figure~\ref{10} shows the variations of polar and dispersive parts of surface tension with the mole fraction of methanol.  It is observed from the results that the strength of intermolecular forces decreases with increase in mole fraction of methanol.  It is also found that, the dispersive forces of ethanol decreases  and hydrogen bond density increases with an addition of methanol to ethanol. It is attributed that methanol molecules are responsible for the hydrogen bonding since ethanol clusters are not strictly hydrogen bonded.  \\ 
\subsection{Molecular dynamics results}
Molecular dynamics simulations were performed to study the nature of interaction between ethanol and methanol molecules of different concentrations, to compare with the experimental results. The simulations were  equilibrated such that they resemble the ambient laboratory conditions.\\

Ethanol aggregates (multimers), ethanol-methanol joint structures, ethanol-methanol-ethanol bridges, ethanol monomers and methanol monomers were commonly observed structures at the various concentrations (Figure~\ref{12}).\\

\begin{figure}[h]
\centerline{\includegraphics[width=5in]{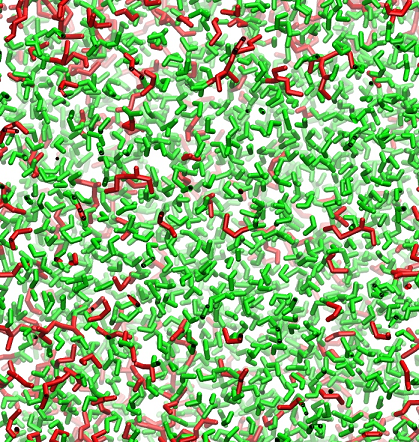}}
\caption[]{As opposed to the lower concentrations, at $80\%$ concentration of methanol in ethanol as shown here, there are fewer number of aggregated structures or clathrates.}%
\label{11}
\end{figure}

\begin{figure}[h]
\centerline{\includegraphics[width=5in]{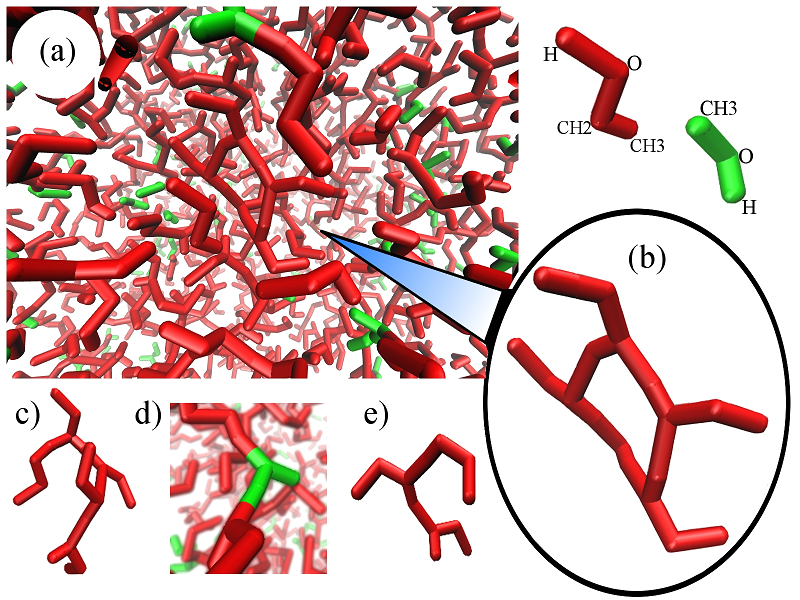}}
\caption[]{The representations of ethanol and methanol as used in the simulations (right top corner)  a) Molecules at $22\%$ concentration of methanol in ethanol b) The ethanol closed cage structure (clathrate) obtained at $22\%$ concentration of methanol in ethanol c) Pentameric open cage structure of ethanol found commonly at several lower concentrations $(<40\%)$ d) A firmly hydrogen bond bridged methanol molecule, holding onto ethanol molecules e) A very common trimeric form of ethanol}%
\label{12}
\end{figure}

\begin{figure}[h]
\centerline{\includegraphics[width=5in]{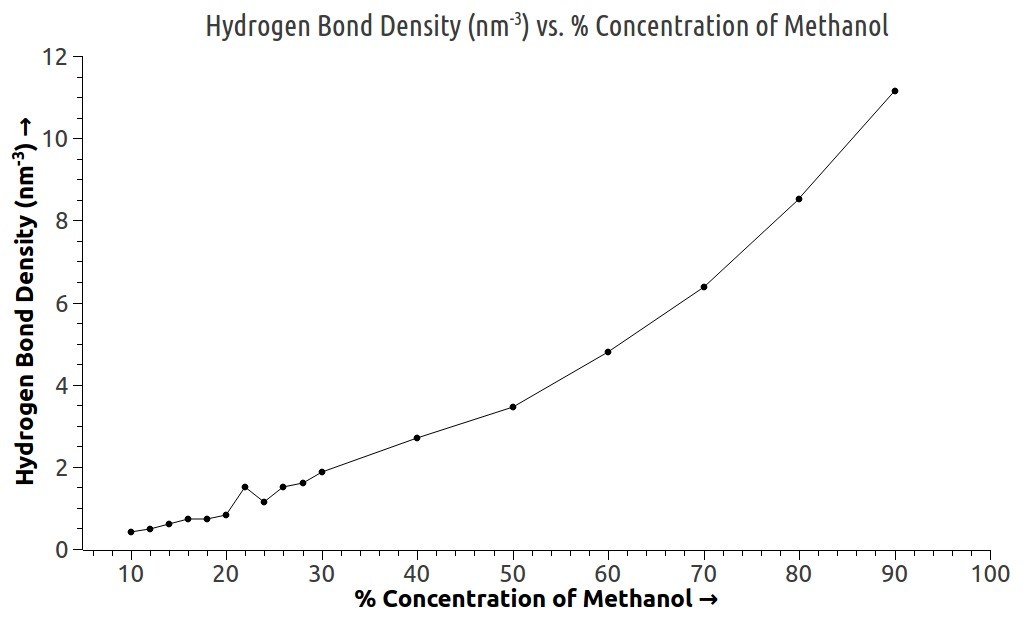}}
\caption[]{Variation of hydrogen bond density with respect to concentration of methanol}%
\label{13}
\end{figure}

Clathrates of various sizes typically dimers to pentamers are observed (Figure~\ref{12}). As the concentration of methanol increases, the number and size of these aggregates reduces and is shown in Figure~\ref{11}. Methanol molecules, with their stronger hydrogen bonding nature insert themselves among the ethanol molecules, sterically hindering ethanol from forming aggregates. This is seen experimentally in Figure~\ref{1} as the regions of constant refractive index. The methanol molecules also act as a bridge between ethanol molecules forming ethanol-methanol-ethanol structures. At concentrations close to $50\%$, ethanol-methanol joint aggregates are also formed in good numbers. At higher concentrations of methanol, ethanol molecules mainly have monomeric forms, while ethanol-methanol joint aggregates became less in number as well.\\
Methanol molecules, however, do not form aggregates among themselves. A graph plotted between hydrogen bonds per volume and percentage concentration of methanol (Figure~\ref{13}) shows that there is a deviation from the expected trend at around $20\%$ concentration. The graph shows that the hydrogen bond density increases with the increase in concentration of methanol. As per the refractive index and surface tension studies, this is the concentration range where complex properties are observed. The formation of these aggregates are proposed to be the reason for these properties.\\
Table 1 provides the output of molecular dynamics simulations.  From the graph it is seen that in the concentration region between $10\%$ and $20\%$, the slope of the graph is much lesser than that at higher concentration regions. In fact in this concentration region the hydrogen bond density increases only by $0.3 nm^{3}$ while at higher concentrations, there are larger changes over a $10\%$ change in concentration. It is clearly understood from this data that there is a direct correlation between the number of molecules of methanol and the H-bond density. The decreased hydrogen bond density at $17\%$ concentration and similar hydrogen bond density at $16\%$ and $18\%$ are also noted from the table. So, the hydrogen bond density, as per the graph, is a function of the number of methanol molecules present, as the ethanol aggregates are not strictly hydrogen-bonded. This is in complete agreement with the experimental results.\\
\begin{table}[h]
\centering
\begin{tabular}{l c c c c c c c}
\hline
Perc. of  & Perc. of & No.of et & No.of met. & No.of  & Box volume  & Hydrogen\\
 ethanol  & methanol &molecules &molecules   & H-bonds &$(nm^{3})$  &bonds/$(nm^{3})$ \\ \hline
90        &10		&1000   &132   &33.71   &80.4956624 &0.41879275\\
88        &12		&1000   &152   &38.834  &77.857765  &0.49878134\\
86        &14		&1000   &185   &50.996  &84.2252341 &0.60547175\\
84        &16		&1000   &200   &62.467  &84.8753878 &0.73598486\\
82        &18   &1000   &226   &64.118  &86.8064943 &0.69175572\\
80        &20   &1000   &254   &74.251  &89.6014556 &0.82868073\\
78        &22   &1000   &283   &97.126  &64.0920729 &1.51541362\\
76        &24   &1000   &310   &108.152 &94.6536029 &1.14260838\\
74        &26   &1000   &350   &148.882 &98.7398907 &1.50782018\\
72        &28   &1000   &386   &162.283 &100.910922 &1.60818073\\
70        &30   &1000   &441   &196.457 &105.081882 &1.86956111\\
60        &40   &1000   &686   &343.271 &127.380453 &2.69484832\\
50        &50   &1000   &1040  &555.289 &160.99895  &3.44902249\\
40        &60   &1000   &1548  &948.363 &198.209334 &4.784655358\\
30        &70   &1000   &2321  &1604.325 &251.864824 &6.3678588\\
20        &80   &1000   &3939  &3066.731 &359.514916 &8.53019128\\
10        &90   &1000   &9072  &7765.461 &696.126002 &11.155252\\ \hline

\end{tabular}
\caption{Variation of hydrogen bond density with the concentration of methanol}
\label{tab:1}
\end{table}

\section{Conclusion}
The surface tension, density, refractive index and dielectric permittivity have been reported for the ethanol-methanol binary system.  The results show distinct type of interactions leading to deviation  in properties at $10\%$ to $30\%$ of methanol. Complex variation in density and surface tension values is observed at $10\%$ to $30\%$ which is attributed to decrease in dipole -dipole interactions and hydrogen bonding that exists in pure ethanol and pure methanol and the increase in intermolecular interaction between ethanol and methanol.  The FTIR results confirm the presence of multimers and tetramers rather than dimers and monomers. Refractive index measurements show the formation of clathrates at $12\%$ to $14\%$ and $21\%$ to $24\%$ of methanol.  The variation in dielectric values at these concentrations clearly indicates the formation of temporarily induced dipoles.  Excess parameters confirmed the domination of dispersion forces at low concentrations in the ethanol-methanol binary system. Molecular dynamics simulations were performed to compare with the experimental results.  It was found from the simulations that the ethanol aggregates are not strictly hydrogen-bond constructed. It can be concluded that methanol involves extensive H-bonding and with increase in methanol concentration, the hydrogen bond density also increases, which agree with the experimental results. 

\section{Acknowledgement}

The authors thank Naval Research Board of India for the Contact Angle Goniometer.


\begin{thebibliography}{100}

\bibitem{markus(1998)}
Markus M. Hoffmann and Mark S. Conradi, "Are There Hydrogen Bonds in Supercritical Methanol and Ethanol?", J. Phys. Chem. B 1998, 102, 263-271

\bibitem{funel(1994)}
Bellissent-Funel, M.-C.; Dore, J. C. "Hydrogen Bond Networks",NATO ASI Series, Series C: Mathematical and Physical Sciences 435;
Kluwer Academic Publishers: Boston, MA, 1994.

\bibitem{saiz(1997)}
L. Saiz, J.A. Padro and E. Guardia, "Structure and Dynamics of Liquid Ethanol", J. Phys. Chem. B 1997, 101, 78-86


\bibitem{wertz(1967)}
D. L. Wertz and R. K. Kruh, "Reinvestigation of the Structures of Ethanol and Methanol at Room Temperature", J. Chem. Phys. 47, 388 (1967)
 
\bibitem{wu(2007)}
Xiaojing Wu , Yuanyuan Chen and Toshio Yamaguchi, "Hydrogen bonding in methanol studied by infrared spectroscopy"', Journal of Molecular Spectroscopy 246 (2007) 187-191

\bibitem{amer(1955)}
H. H. Amer and A. R. Paxton, "Methanol-ethanol-acetone-Vapour liquid equilibria", Industrial and engineering chemistry, Vol.48, No. 1, 142-146, june 1955. 

\bibitem{per(1969)}
Per Dalager, "Vapor-liquid Equilibria of Binary Systems of Water with Methanol and Ethanol at Extreme Dilution of the Alcohols", J. Chem. Eng. Data., Vol. 14, No. 3, July 1969

\bibitem{xiao(1997)}
Caibin Xiao, Hugo Bianchi, and Peter R Tremaine,"Excess molar volumes and densities of methanol-water at temperatures between 212 K and 462 K and pressures of 7 MPa and 13.5 MPa",J. Chem. Thermodynamics, 1997, 29, 261-286. 

\bibitem{jose(2015)}
Jose J, Cano-Gomez, Gustavo A, Iglesias-Silva, Edgar O, Castrejon-Gonzalez, Mariana Ramos-Estrada, and Kenneth R. Hall, "Density and Visocsity of Binary liquid mixtures of ethanol and 1-hexanol and ethanol+1-heptanol from 293.15 to 328.15 K at 0.1 MPa", J. Chem. Eng. Data. 2015, 60, 1945-1955 

\bibitem{lone(2011)}
Baliram Lone and Vinjanmpaty Madhurima, " Dilectric and conformal studies of 1-propanol and 1-butanol in methanol", J. Mol. Model(2011) 17:709-719

\bibitem{ezekial(2012)}
Ezekiel D. Dikio1, Simphiwe M. Nelana, David A. Isabirye, Eno E. Ebenso, "Density, Dynamic Viscosity and Derived Properties of Binary Mixtures of Methanol, Ethanol, n-Propanol, and n-Butanol with Pyridine at T $=$ (293.15, 303.15, 313.15 and 323.15) K", Int. J. Electrochem. Sci., 7 (2012) 11101 - 11122

\bibitem{hanwell(2012)}
Marcus D Hanwell, Donald E Curtis, David C Lonie, Tim Vandermeersch, Eva Zurek and Geoffrey R Hutchison; "Avogadro: An advanced semantic chemical editor, visualization, and analysis platform" Journal of Cheminformatics 2012 , 4:17

\bibitem{schmidt(1993)}
 M.W.Schmidt, K.K.Baldridge, J.A.Boatz, S.T.Elbert, M.S.Gordon, J.H.Jensen, S.Koseki, N.Matsunaga, K.A.Nguyen, S.Su, T.L.Windus, M.Dupuis, J.A.Montgomery, General Atomic and Molecular Electronic Structure System,  J. Comput. Chem. , 14, 1347-1363(1993)

\bibitem{root(1951)}
C.C.J.Roothaan, New Developments in Molecular Orbital Theory, Rev.Mod.Phys.  23, 69-89(1951)

\bibitem{szabo(1996)}
Szabo, A.; Ostlund, N. S. (1996). Modern Quantum Chemistry. Mineola, New York: Dover Publishing. ISBN 0-486-69186-1.

\bibitem{hari(1973)}
P.C.Hariharan, J.A.Pople; The influence of polarization functions on molecular orbital hydrogenation energies,  Theoret.Chim.Acta 28, 213-222(1973)

\bibitem{franc(1982)}
M. M. Francl, W. J. Pietro, W. J. Hehre, J.S. Binkley, M.S. Gordon, D.J. De Frees, J. A. Polpe, Self-consistent molecular orbital methods. XXIII, A Polarization-type basis set for second-row elements, J. Chem. Phys. 77, 3654-3665 (1982)


\bibitem{becker(1993)}
Bekker, H., Berendsen, H. J. C., Dijkstra, E. J., Achterop, S., van Drunen, R., van der Spoel, D., Sijbers, A., Keegstra, H., Reitsma, B., Renardus, M. K. R. Gromacs: A parallel computer for molecular dynamics simulations. In Physics Computing 92 (Singapore, 1993).
de Groot, R. A., Nadrchal, J., eds. . World Scientific.

\bibitem{berendsen(1995)}
Berendsen, H. J. C., van der Spoel, D., van Drunen, R. GROMACS: A message-passing parallel molecular dynamics implementation. Comp. Phys. Comm. 91 : 43-56, 1995

\bibitem{lindahl(2001)}
Lindahl, E., Hess, B., van der Spoel, D. GROMACS 3.0 "A package for molecular simulation and trajectory analysis". J. Mol. Mod. 7:306-317, 2001.

\bibitem{van(2005)}
Van der Spoel, D., Lindahl, E., Hess, B., Groenhof, G., Mark, A. E., Berendsen, H. J. C. GROMACS: Fast, Flexible and Free. J. Comp. Chem. 26:1701-1718, 2005.

\bibitem{hess(2008)}
Hess, B., Kutzner, C., van der Spoel, D., Lindahl, E. GROMACS 4: Algorithms for Highly Efficient, Load-Balanced, and Scalable Molecular Simulation. J. Chem. Theory Comput. 4(3) : 435-447, 2008.

\bibitem{pronk(2013)}
Pronk, S., Pall, S., Schulz, R., Larsson, P., Bjelkmar, P., Apostolov, R., Shirts, M. R., Smith,
J. C., Kasson, P. M., van der Spoel, D., Hess, B., Lindahl, E. GROMACS 4.5: a high-throughput and highly parallel open source molecular simulation toolkit. Bioinformatics 29(7): 845-854, 2013.

\bibitem{pall(2015)}
 Pall, S., Abraham, M. J., Kutzner, C., Hess, B., Lindahl, E. Tackling exascale software challenges in molecular dynamics simulati ons with GROMACS. In: Solving Software Challenges for Exascale. Vol. 8759. Markidis, S., Laure, E. eds. Vol.8759. . Springer International Publishing Switzerland London 2015 3-27.

\bibitem{abraham(2015)}
Abraham, M. J., T., Schulz, R., R., Pall, S., Smith, J. C., Hess, B., Lindahl, E. GROMACS: High performance molecular simulations through multi-level parallelism from laptops to supercomputers. SoftwareX 1-2, 19-25, 2015.

\bibitem{humphrey(1996)}
Humphrey, W., Dalke, A. and Schulten, K., "VMD - Visual Molecular Dynamics", J. Molec. Graphics, 1996, vol. 14, pp. 33-38.


\bibitem{zimmer(1991)}
Zimmerman, K., All purpose molecular mechanics simulator and energy minimizer,  J. Comp. Chem. 12:310-319, 1991.

\bibitem{parri(1981)}
 M. Parrinello, A. Rahman, Polymorphic transitions in single crystals: A new molecular dynamics method, J . Appl. Phys . 52 ( 1981 ) 7182

\bibitem{beren(1984)}
Berendsen, H. J. C. ; Postma, J. P. M.; van Gunsteren, W. F.; DiNola, A.; Haak, J. R.  "Molecular-Dynamics with Coupling to an External Bath". Journal of Chemical Physics 81(8): 3684-3690, 1984.

\bibitem{sara(2012)}
Saravanakumar K and Kubendran T.R., Density and Viscosities for the Binary Mixtures of 1, 4-Dioxane and Benzene
or Chlorobenzene at 303.15, 308.15, 313.15 K and a Pressure of 0.1MPa, Research Journal of Chemical Sciences, Vol. 2(4), 50-56, April (2012)

\end{thebibliography}

\end{document}